| | |
|---|---|
| Abstract | $CO_2$ conversion into value-added products has gained significant interest over the few last years, as the greenhouse gas concentrations constantly increase due to anthropogenic activities. Here we report on experiments for $CO_2$ conversion by means of a cold atmospheric plasma using a cylindrical flowing dielectric barrier discharge (DBD) reactor. A detailed comparison of this DBD ignited in a so-called burst mode (i.e. where an AC voltage is applied during a limited amount of time) and pure AC mode is carried out to evaluate their effect on the conversion of $CO_2$ as well as on the energy efficiency. Decreasing the duty cycle in the burst mode from 100% (i.e. corresponding to pure AC mode) to 40% leads to a rise in the conversion from 16–26% and to a rise in the energy efficiency from 15 to 23%. Based on a detailed electrical analysis, we show that the conversion correlates with the features of the microfilaments. Moreover, the root-mean-square voltage in the burst mode remains constant as a function of the process time for the duty cycles <70%, while a higher duty cycle or the usual pure AC mode leads to a clear voltage decay by more than 500 V, over approximately 90 s, before reaching a steady state regime. The higher plasma voltage in the burst mode yields a higher electric field. This causes the increasing the electron energy, and therefore their involvement in the $CO_2$ dissociation process, which is an additional explanation for the higher $CO_2$ conversion and energy efficiency in the burst mode. |

# Introduction

New records in greenhouse gas concentrations are reported every year. In particular carbon dioxide ($CO_2$) is considered an inevitable reaction product of human activity. Indeed, the anthropogenic activities have led to increasing atmospheric $CO_2$ concentrations over the last 150 years, from 280 to almost 400 ppm in 2014 [1]. This means higher global temperatures and more extreme weather phenomena such as: heat waves, floods, ice melting, rising of sea levels, and ocean acidity [2–7]. This concern has recently been reported during the three-week-long COP-21 conference in Paris to prompt governments to take stronger measures in order to limit global warming. Nowadays, the political authorities want to limit the production of $CO_2$ by promoting renewable energies, noncarbon energies and geological storage in deep underground [8–10] or by taxing the tons of $CO_2$ produced. As an innovative alternative, there is an increasing interest in the possibility to reutilize $CO_2$, i.e. in a 'cradle to cradle' approach, where waste of all types is being recycled. Carbon dioxide—the ultimate gaseous combustion product of carbon compounds— could be reutilized, although it is not straightforward, because it is a particularly stable molecule ($\Delta G_f \ll 0$).

In recent years, many publications have reported on $CO_2$ recycling following conventional organic reactions [11], electrochemical reduction in solutions [12] or reduction in plasma phase [13]. The latter technique appears very promising, as it can directly operate in the gaseous phase, whilst electrochemistry techniques must be carried out in the liquid phase. Moreover, if it is mixed with hydrogen, water or methane, $CO_2$ can easily be reduced into syngas ($CO/H_2$) and small organic molecules [14–21]. The synthesis of value added compounds from $CO_2$, considered as a 'raw material', is a very interesting concept for the industrial sector. However, as $CO_2$ corresponds to a particularly stable oxidized state of carbon, its reduction is not spontaneous and requires significant energy input. In cold atmospheric plasmas, such as dielectric barrier discharges (DBD), this energy is provided by electric power. The latter mainly heats the electrons, while the gas itself can remain near room temperature, thus allowing the reaction to procede at mild reaction conditions (ambient pressure and temperature). Indeed, the energetic electrons will activate the gas molecules by excitation, ionization and dissociation reactions, creating reactive species that can easily form new molecules.

A DBD is commonly supplied with an AC voltage (pure AC mode) (e.g. [22–24]) and sometimes by high voltage nano or micro pulses [25–35] (pulsing mode). According to Jiang et al [25], supplying a DBD with high voltage nanopulses enhances its stability over time, as well as its uniformity, since no irregular distribution of microdischarges is revealed. Such a power delivery presents great benefits for plasma polymerization, thin layer deposition and surface treatments as no significant texturization appears and it allows, for instance, an optimal deposition of a subsequent barrier layer [36–38]. This configuration is also of interest for gas treatment since a larger number of microdischarges are produced, increasing the probability for a single gas molecule to pass through the discharge and interact with a single microdischarge. Ndong et al explain that positive voltage pulses create discrete discharge channels uniformly arranged on the dielectric surface in a streamer-like regime, while negative voltage pulses mostly tend to generate a uniform discharge along the active electrode such as a glow-like discharge [39, 40]. Song et al investigated a bipolar pulsed discharge for the $CO_2$ reforming of methane where both the $CO_2$ and $CH_4$ conversion are higher when applying a pulse mode compared to a classical sinusoidal AC mode [41]. Some other authors have also compared a unipolar and bipolar DC discharge [42–44] and showed that the $CO_2$ conversion slightly increases upon rising pulse frequency.

Besides the usual pure AC and pulsing modes, a DBD can also be supplied by an AC voltage that is switched off at regular times, at a frequency much lower than the AC frequency. This so-called burst mode is well known in manufacturers' manuals and also in the literature dedicated to power supplies [45–53]. Benard et al [45] have demonstrated how the production of small perturbations to the classical alternative voltage (i.e. burst mode) can enhance the stability of the DBD. However, to the authors' knowledge, there exist no papers yet which deal with cold atmospheric plasma sources operating in this so-called burst mode for gas conversion.

In the present article, we investigate the splitting of $CO_2$ using a flowing DBD operating in burst mode and we compare its performance with the usual pure AC mode. This burst mode corresponds to a sinusoidal voltage with a signal frequency of 28.6 kHz (period of 35.7 µs), which is switched on and off at a repetition frequency between 400 and 900 Hz. The influence of this mode is investigated on the $CO_2$ conversion, the energy efficiency and the microdischarge features (i.e. number, lifetime and charge of the microdischarges) and we compare the results with the same DBD that is supplied with a pure AC voltage (duty cycle = 100%). Electrical characterization, mass spectrometry and optical emission spectroscopy are applied to understand how and why the CO2 conversion is enhanced.

# Experimental set-up

## DBD reactor and voltage measurements

The $CO_2$ gas flows through a cylindrical DBD reactor as shown in figure 1. The $CO_2$ flow rate is set at 200 $mL_n \cdot min^{-1}$ and the operating frequency is fixed at 28.6 kHz. The central cylindrical copper electrode has a diameter of 22 mm and a length of 120 mm. This inner electrode is powered by an AC high voltage, whereas the outer electrode is grounded. The latter is a mesh, made of stainless steel, with length of 100 mm (thus defining the discharge length), and placed around a dielectric barrier (2 mm in thickness), made of alumina and with inner diameter of 26 mm, yielding a discharge gap of 2 mm, and thus a discharge volume of 15.1 cm3. This corresponds to a residence time of the gas equal to 4.5 s for the given gas flow rate of 200 $mL_n \cdot min^{-1}$. The applied power is provided by an AFS generator G10S-V and a transformer, which amplify the sinusoidal signal of the alternating voltage and which allow us to pulse the voltage in a controlled manner.

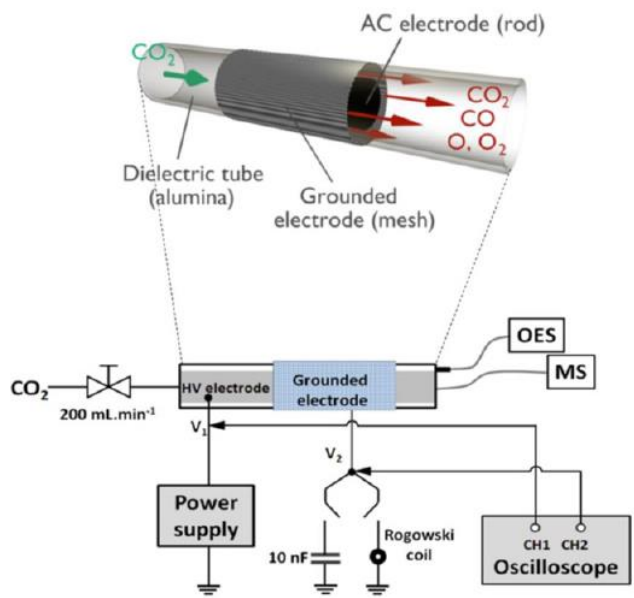

*Figure 1. Schematic diagram of the experimental set-up.*

In the present study, the signal frequency remains fixed at $f_{signal}$ = 28.6 kHz (period of 35.7 μs). Furthermore, $T_{ON}$ and $T_{OFF}$ are defined as the time of plasma ignition and the time when no power is applied to the reactor. Except for the analysis of $T_{ON}$ and $T_{OFF}$ at constant duty cycle, the main results described in this article are related to the influence of the duty cycle. In this latter case, $T_{ON}$ remains fixed at 1 ms while $T_{OFF}$ is changed between 0 and 1.5 ms, corresponding to a duty cycle ($D_{cycle}$) between 100 and 40%. Therefore, the repetition frequency of the burst mode, $f_{repetition}$ = $1/(T_{ON} + T_{OFF})$, is varied between 400 and 900 Hz. The voltage as a function of time is plotted in figure 2 for high voltages delivered either in a pure AC mode (a) or in a burst mode (b). As the total power applied to the plasma ($P_{applied}$) is fixed at 50 W in all cases, the power during the plasma ignition ($P_{plasma}$) in burst mode is always higher than in pure AC mode, since $P_{plasma}$ (W) = $P_{applied}$ (W)/$D_{cycle}$ (%).

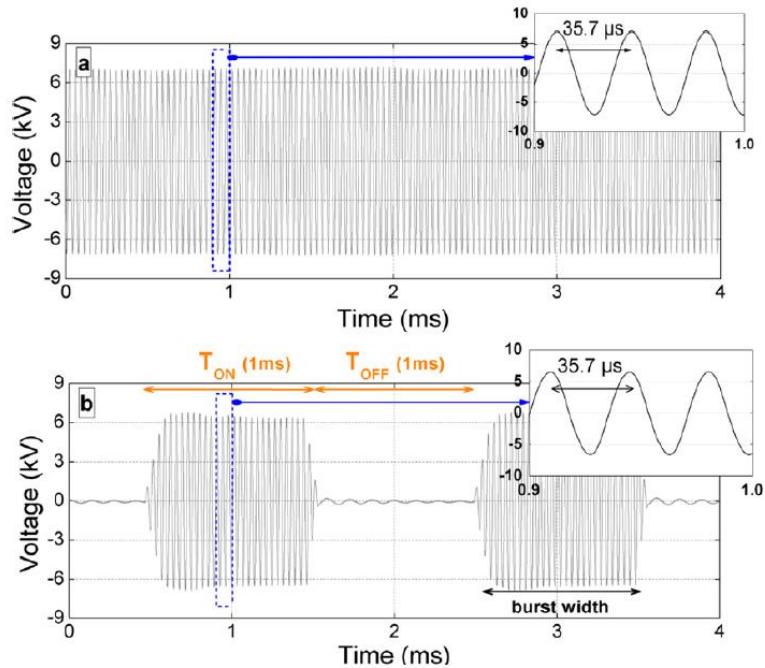

*Figure 2. High voltage signals applied to the discharge, with frequency of 28.6 kHz (or period of 35.7 μs), in (a) pure AC mode, and (b) burst mode, with a duty cycle of 50% ($T_{ON} = T_{OFF}$ = 1 ms).*

## Mass spectrometry (MS)

The gaseous products from the post-discharge are analyzed by a mass spectrometer operating at atmospheric pressure (Hiden analytical QGA). The electron energy in the ionization chamber is set at 70 eV and a secondary electron multiplier (SEM) is used as a detector. The $CO_2$ conversion ($CO_2$) is calculated according to the following equation, where I stands for the intensity of the signals:

$$\chi_{CO_2}(\%) = \frac{I_{CO_2 \text{ without plasma}} - I_{CO_2 \text{ with plasma}}}{I_{CO_2 \text{ without plasma}}} \times 100\%$$

The energy efficiency of the $CO_2$ conversion is calculated from $CO_2$, the enthalpy of the splitting reaction ($CO_2 \rightarrow CO + \frac{1}{2}O_2$), namely $H°_{298K}$ = 279.8 kJ · mol$^{-1}$ = 2.9 eV · mol$^{-1}$ and the specific energy input (SEI), according to the following equation:

$$\eta_{CO_2}(\%) = \chi_{CO_2}(\%) \cdot \frac{\Delta H^0_{298K}(eV \cdot mol^{-1})}{SEI\ (eV \cdot mol^{-1})}$$

The SEI is defined as follows:

$$SEI\ (eV \cdot mol^{-1}) = \frac{\frac{P_{applied}\ (J \cdot s^{-1})}{Gas\ flow\ rate\ (cm^3 \cdot s^{-1})} \times 6.24 \times 10^{18}(eV \cdot J^{-1}) \times 24\,500\ (cm^3 \cdot mol^{-1})}{6.022 \times 10^{23}(molecule \cdot mol^{-1})}$$

In principle, GC analysis is more accurate for determining the $CO_2$ conversion, but it was not available for this study. We believe that MS analysis is also sufficiently accurate at the current conditions.

## Electrical measurements

A Tektronix DPO 3032 oscilloscope is used to perform electrical measurements (during $T_{ON}$). The first channel is dedicated to the monitoring of the high voltage ($V_1$) with a Tektronix P6015A probe, whilst the second channel allows the measuring of the voltage passing through a capacitor, to obtain the power absorbed by the plasma via the Lissajous method [54–58], or through a Rogowski coil (Pearson 2877) to probe the current. The capacitor or the Rogowski coil is placed in series with the DBD. The obtained current profiles as a function of time, also called oscillograms, correspond to 2 different types of current: the plasma current and the dielectric current. These two can be simply distinguished as the dielectric component is a sinusoidal-like signal, whereas the plasma component corresponds to hundreds of microdischarges occurring at each half period, since an atmospheric pressure DBD operating in $CO_2$ generally operates in a filamentary regime [24, 59–61]. Analysis of these microdischarges is important to better understand the $CO_2$ splitting process. A statistical study will be carried out for several periods in order to evaluate the magnitude of different electrical parameters of these microdischarges. By applying a numerical method described in our previous article [61], the number of microdischarges ($N_{md}$), their lifetime ($L_{md}$), the plasma charge accumulation ($Q_{pl}$) and the resulting plasma current ($i_{pl}$) can be achieved. These data are collected for several periods and then averaged over a single period in order to have statistically valid results. In particular, $N_{md}$ has been measured during TON = 1 ms and multiplied by $D_{cycle}$ in order to consider the same residence time. Likewise, the applied voltage (VDBD) can also be expressed as the sum of two voltages: the dielectric voltage ($V_{diel}$) and the plasma voltage ($V_{pl}$), which typically represents 80% of the total VDBD [61, 62].

## Optical emission spectroscopy (OES)

The OES measurements are performed with an Andor Shamrock-500i spectrometer including an Andor DU420A-OE CCD camera. Each spectrum is acquired for an exposure time of 5 s and 7 accumulations. Of particular interest is the CO emission band headed at 483.3 nm from the CO Angstrom system (B1$\Sigma^+$ → A$^1\Pi$; v'=0 → v''=1) shown in figure 3. Its ro-vibrational structure is calculated and fitted to the experimental spectra in order to determine the rotational temperature ($T_{rot}$) and hence the gas temperature ($T_{gas}$), assuming $T_{rot} \approx T_{gas}$ [63–65]. For the sake of comparison, the gas temperature is also measured using another band: the first positive system (FPS) of $N_2$ is analyzed by adding 10% of nitrogen to $CO_2$. In this case, the temperature is measured via a line-ratio peak formula, specific to this system [66, 67], using the peaks at 775.3 and at 773.9 nm.

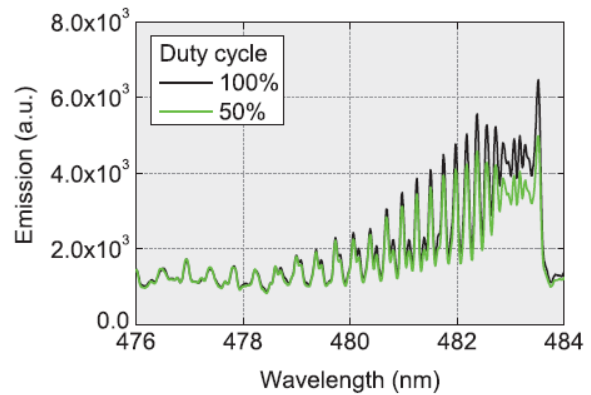

Figure 3. CO band from the [$B_1\Sigma^+$ → $A^1\Pi$; v' = 0 → v'' = 1] Angstrom System in the DBD supplied with pure $CO_2$, at two different duty cycles.

# Results and discussion

## $CO_2$ conversion and energy efficiency

The $CO_2$ conversion and energy efficiency as a function of the duty cycle of the burst mode are depicted in figure 4(a). The conversion obtained in the pure AC mode ($D_{cycle}$ = 100%) is 16.1%, which is significantly lower than in the burst mode at 40% duty cycle (i.e. 25.8%). Decreasing the duty cycle at constant applied power thus results in a higher CO2 splitting. As evidenced in figure 4(b), smaller duty cycles yield higher peak powers, e.g. 100 W for $D_{cycle}$ = 50% versus 50 W for $D_{cycle}$ = 100%. These higher peak powers delivered within shorter times can explain the higher $CO_2$ conversion. However, we will also show at the end of the paper that the enhancement of the burst mode is not only attributed to the higher peak power, but also to other effects (see below). The corresponding energy efficiency also increases upon decreasing duty cycle from 14.5% (at $D_{cycle}$ = 100%) to 23.1% (at $D_{cycle}$ = 40%).

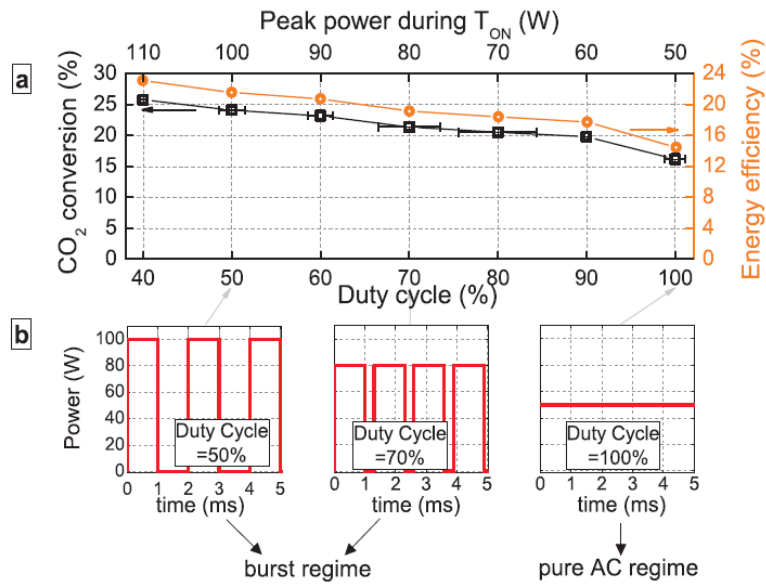

*Figure 4. (a) $CO_2$ conversion and energy efficiency as a function of the duty cycle — $P_{applied}$ = 50 W, $f_{signal}$ = 28.6 kHz, $f_{repetition}$ = 400–900 Hz and $\Phi(CO_2)$ = 200 $mL_n \cdot min^{-1}$; (b) power versus time for $D_{cycle}$ = 50, 70 and 100%.*

On the other hand, keeping the duty cycle constant but increasing the burst width from 1 ms to even more than 20 ms has no effect on the $CO_2$ conversion and energy efficiency, as illustrated in figure 5. Therefore, in the next sections we will only investigate the influence of $T_{OFF}$ (or duty cycle or repetition frequency) on the electrical properties of the discharge, and more specifically on the microdischarge features. First we will show the effect of the duty cycle on the evolution of the applied voltage with increasing process time, to check the time-stability of the discharge. Next, the influence of the duty cycle on the current oscillograms and on the microdischarge properties, as well as on the gas temperature will be studied. Finally, the burst mode and the normal AC mode will be compared at the same delivered power, to illustrate that the enhancement of the burst mode is not only attributed to the higher peak power, but also due to the microdischarge properties.

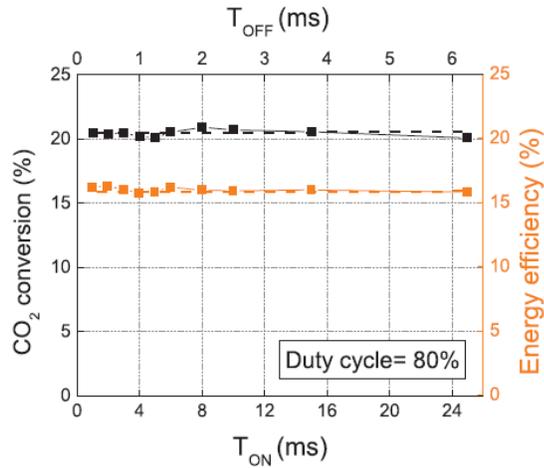

*Figure 5. $CO_2$ conversion and energy efficiency for a fixed duty cycle of 80%, as a function of $T_{ON}$. $P_{applied}$ = 50 W, $f_{signal}$ = 28.6 kHz and $\Phi(CO_2)$ = 200 $mL_n \cdot min^{-1}$.*

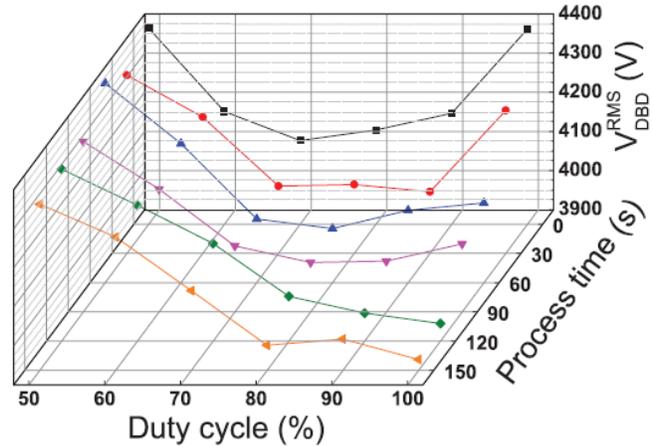

*Figure 6. RMS value of the DBD voltage as a function of the duty cycle and the process time. $P_{applied}$ = 50 W, $f_{signal}$ = 28.6 kHz, $f_{repetition}$ = 500–900 Hz and $\Phi(CO_2)$ = 200 $mL_n \cdot min^{-1}$.*

## Influence of the duty cycle on the discharge time-stability

Figure 6 shows the applied voltage as a function of the process time for several different duty cycles. The lower the duty cycle, the more stable and the higher the voltage is as a function of the process time. For $D_{cycle}$ < 70%, the voltage remains quite stable over time, e.g. at $D_{cycle}$ = 50% it is close to 4400 V for all process times investigated (with only a small variation within the first 60 s). Therefore, the discharge appears as very stable in time. However, this time-stability gradually vanishes upon increasing Dcycle and in the pure AC mode ($D_{cycle}$ = 100%) the time dependency is quite significant: a drop of approximately 500 V is reached in 90 s, from 4400 V to a plateau at 3900 V. This transient regime is clearly visible in the pure AC mode and is due to the dielectric properties of the barrier: it behaves as a capacitor with a dielectric relaxation time that is much longer than the period associated with the applied voltage. The dielectric relaxation time of the barrier trelax is defined as the product of its dielectric permittivity ($\varepsilon = \varepsilon_0 \cdot \varepsilon_r$) with its electrical resistivity ($\rho$), i.e. $t_{relax} = \varepsilon_0 \cdot \varepsilon_r \cdot \rho$, where $\varepsilon_0$ = 8.854 · $10^{-12}$ F·$m^{-1}$ and, in the case of alumina, $\varepsilon_r$ = 9 and $\rho \approx 10^{14}$ $\Omega^{-1}$·$cm^{-1}$ [68]. A rough estimation leads to trelax ≈ 0.8 s. This value is much higher than the period of the applied voltage, which is 35.7 µs (since fsignal = 28.6 kHz). In the first half period, the electrical charge deposited on the barrier cannot be entirely evacuated and remains partially in the second half period. This charge accumulation is negligible at the very beginning of the discharge ignition, but becomes significant in less than 30 s of operation (see figure 6 for $D_{cycle}$ = 100%). Once the DBD is ignited, the barrier potential gradually increases, hence inducing a gradual decrease in $V_{DBD}$ observed in figure 6. This phenomenon is clearly less pronounced in the burst mode, where the $T_{ON}$ times are too short to induce significant charge accumulations on the barrier. Moreover, the $T_{OFF}$ times are much longer so that the sustaining voltage can remain closer to the ignition voltage [69]. The applied voltage ($V_{DBD}$) is expressed as the sum of two voltages, $V_{DBD} = V_{pl,eff} + V_{diel}$, where $V_{pl,eff}$ is the effective plasma voltage (in the gap) and $V_{diel}$ is the voltage through the dielectric barrier. In a previous work [61], we have demonstrated that varying $V_{DBD}$ for a fixed power induces variations of $V_{pl,eff}$ while $V_{diel}$ remains almost unchanged. Here, decreasing the duty cycle at $P_{applied}$ = 50 W only induces an increase in $V_{DBD}$, hence of $V_{pl,eff}$. As the gap is fixed, we can thus conclude that a stronger electric field is created as the plasma voltage is higher, and therefore there will be more energetic electrons. As a result, the burst mode can enhance the rates of electron impact excitation, ionization and dissociation of $CO_2$ and thus improve the conversion and energy efficiency of the $CO_2$ splitting process, as illustrated in figure 4(a). Hence, besides the higher peak power (due to the shorter duty cycle), this can also explain the higher $CO_2$ conversion in the burst mode.

## Influence of the duty cycle on the gas temperature

The influence of the duty cycle on the gas temperature is obtained from OES measurements. Ro-vibrational spectra of the Angstrom CO band are investigated, as well as the first positive system (FPS) of $N_2$. In figure 7, a linear decrease in gas temperature is observed when reducing the duty cycle, i.e. when increasing the $T_{OFF}$ times. This is logical because in the burst mode, the gas can cool down. In the pure AC mode, where the $CO_2$ conversion is the lowest (16.1%), the gas temperature is the most elevated, suggesting that in general, a larger fraction of the power is used for gas heating, while a smaller fraction of the power leads effectively to $CO_2$ splitting. It should be noted that the gas temperature as obtained from the FPS of $N_2$ is about 100 K higher than the gas temperature obtained from the Angstrom CO band, over the entire range of duty cycles investigated. At this point, it is rather difficult to explain this difference since no other independent method (e.g. Doppler broadening, thermocouple, etc.) was used for the gas temperature determination in the plasma phase. Besides this, it is also worth mentioning that the line-ratio and synthetic spectra calculation methods are prone to errors as they assume a Boltzmann distribution and the validity of this assumption cannot be directly checked. Nevertheless, despite the difference of about 100 K, the trend of decreasing gas temperature with decreasing duty cycle is obvious from both curves, as shown in figure 7.

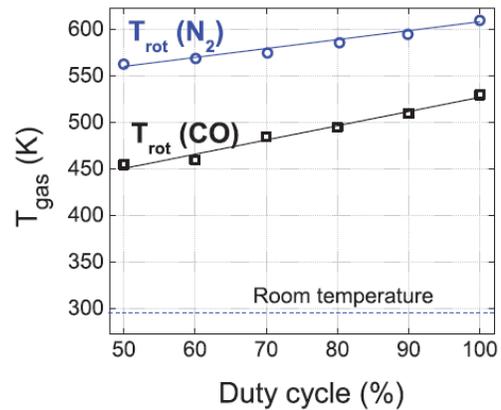

*Figure 7. Gas temperature as a function of duty cycle, as obtained from the FPS of $N_2$ and from the Angstrom CO band; $P_{applied}$ = 50 W, $f_{signal}$ = 28.6 kHz, $f_{repetition}$ = 500–900 Hz and $\Phi(CO_2)$ = 200 $mL_n \cdot min^{-1}$.*

## Influence of the duty cycle on the electrical characteristics

Figure 8(a) shows a picture of the filaments in the DBD for two different duty cycles (50 and 100%). For this purpose, the alumina dielectric barrier is replaced by (transparent) borosilicate glass in order to visually observe these filaments. The pictures reveal an apparent larger number of filaments at $D_{cycle}$ = 50% than in pure AC mode ($D_{cycle}$ = 100%). To study this in more detail, these pictures are correlated with their corresponding current oscillograms in figure 8(b), illustrating the currents peaks ($I_{DBD}$ = $I_{pl}$ + $I_{diel}$) during one AC period, i.e. 35.7 μs. Note that the oscillogram corresponding to $D_{cycle}$ = 50% is a current profile related to $P_{plasma}$ = 100 W, as these oscillograms are recorded only during $T_{ON}$. A visual observation of these oscillograms indicates that the number of microdischarge pulses, as well as their lifetime and the current amplitude all increase from $D_{cycle}$ = 100% to $D_{cycle}$ = 50%. Based on a numerical method explained in [61], we can also obtain more quantitative information on these microdischarges, such as their number ($N_{md}$), lifetime ($L_{md}$), charge ($Q_{pl}$) and conduction current ($i_{pl}$), from these current peaks.

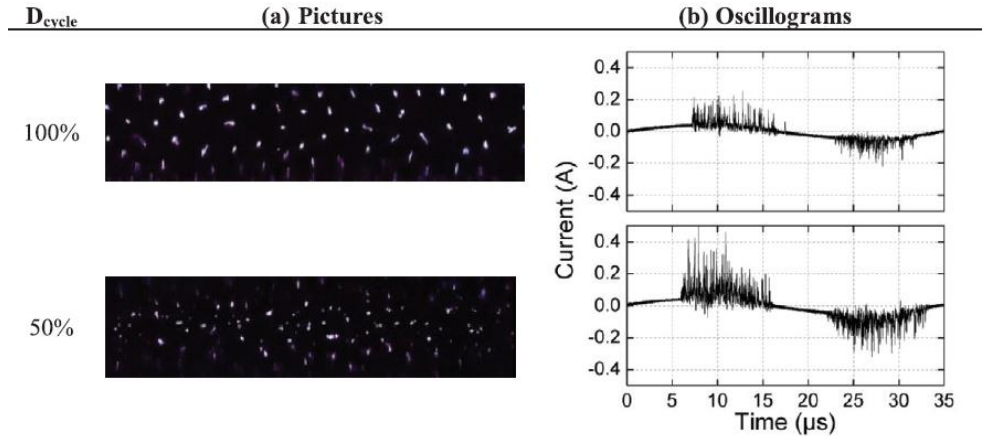

Figure 8. (a) Pictures of the distribution of the microdischarges in the DBD, and (b) corresponding current oscillograms, for two different duty cycles (50 and 100%). $P_{applied}$ = 50 W, $f_{signal}$ = 28.6 kHz and $\Phi(CO_2)$ = 200 mL$_n$ · min$^{-1}$. Camera aperture is 1/100 s.

These parameters are plotted as a function of duty cycle in figure 9, as obtained for one AC period of 35.7 µs. These results refer to the same applied power, and thus the peak power is higher in the burst mode. All these parameters are given in terms of (i) the same residence time (by taking into account $T_{OFF}$ and thus, the electrical parameters are multiplied by $D_{cycle}$) and (ii) only the plasma ignition (without taking into account $T_{OFF}$). As observed in the figures, all of them are clearly changed in the burst mode and upon decreasing duty cycle. For instance in terms of same residence time, a reduction of $D_{cycle}$ from 100 to 50% induces a decrease in the $N_{md}$ from 390 to 340 per period. On the other hand, their mean lifetime ($L_{md}$) increases from 12.9–14.8 ns during the plasma ignition. A drastic rise of the plasma current $i_{pl}$ (from 9–14 mA) is also observed when the duty cycle drops from 100 to 50%, which will result in a significant increase in electron density. The latter can be deduced from the rise in plasma charge $Q_{pl}$ in terms of a given residence time (from 0.3–0.5 µC). As the oscillograms are recorded during $T_{ON}$, and the peak power becomes higher in the burst mode, the values of $N_{md}$, $L_{md}$, $Q_{pl}$ and $i_{pl}$ will increase upon increasing plasma power $P_{plasma}$ (see our previous study on the effect of power on the electrical characteristics [61]). Therefore in the last paragraph, we will also compare the electrical characteristics in burst mode and pure AC mode, at the same delivered power during $T_{ON}$.

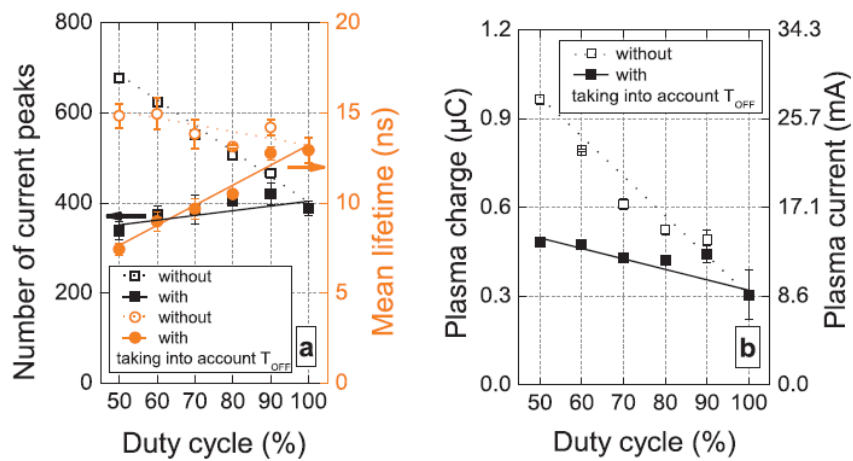

Figure 9. (a) Number of microdischarge over one period (left axis) and their mean lifetime (right axis) versus duty cycle, and (b) plasma charge accumulation over one period (left axis) and plasma current (right axis) versus duty cycle. All electrical parameters are given either considering a total residence time (solid lines) or $T_{ON}$ (dot lines). $P_{applied}$ = 50 W, $f_{signal}$ = 28.6 kHz, $f_{repetition}$ = 500–900 Hz and $\Phi(CO_2)$ = 200 mL$_n$ · min$^{-1}$.

## Correlation between microdischarge parameters and energy efficiency of burst and pure AC mode

Figure 10 depicts the number of microdischarges, their mean lifetime and the plasma charge and current for the pure AC (values adopted from our previous paper [61]) and burst mode, as a function of plasma power. Several observations are noteworthy:

(i) In pure AC mode (i.e. all corresponding to $D_{cycle}$ = 100%), the plasma power corresponds to the applied power ($P_{plasma} = P_{applied}$). Thus, the plasma power is increased by tuning the applied power.

(ii) The plasma power in burst mode corresponds to $P_{plasma} = P_{applied}/D_{cycle}$ where $P_{applied}$ is fixed (50 W). The plasma power (delivered during $T_{ON}$) is increased by tuning the duty cycle. For example, $P_{plasma}$ = 50 W at $D_{cycle}$ = 100% and ca. 100 W at $D_{cycle}$ = 50%. The corresponding duty cycles are indicated next to the black points in the figure.

(iii) All the aforementioned results in the article can only be compared to the black curves in figure 10, i.e. for $P_{applied}$ = 50 W

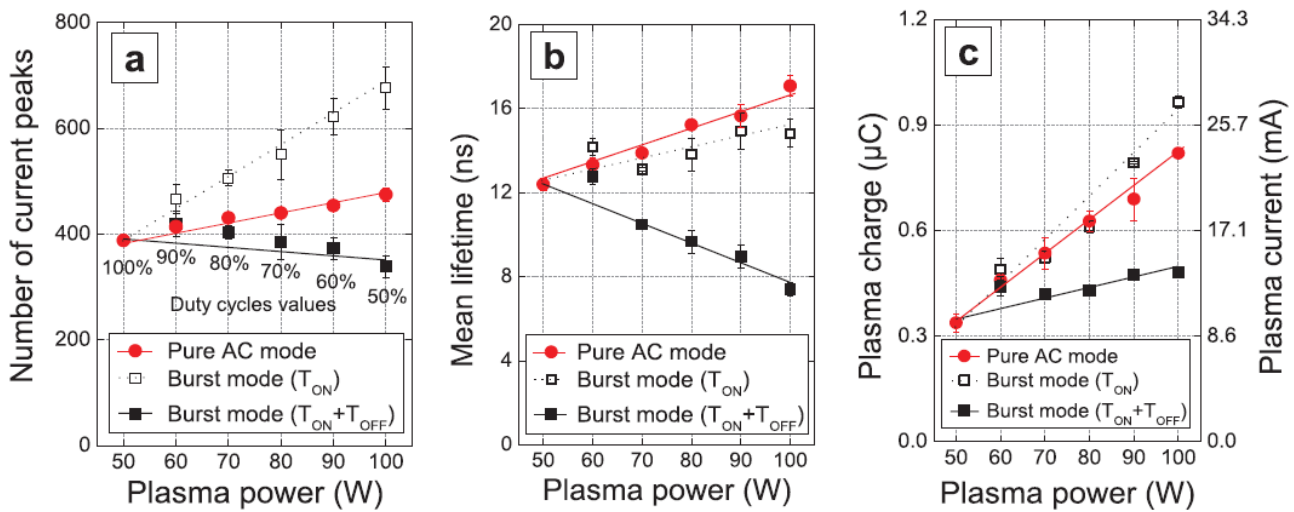

*Figure 10. (a) Number of microdischarges during one period, (b) microdischarges mean lifetime, and (c) plasma charge accumulation/ plasma current during one period, as a function of the plasma power. $P_{applied}$ is fixed in the burst mode while $D_{cycle}$ is varied. $P_{applied}$ is changed in the pure AC mode (electrical parameter values are adopted from our previous article [61]). $f_{signal}$ = 28.6 kHz and $\Phi(CO_2)$ = 200 $mL_n \cdot min^{-1}$.*

In figure 10(a), the number of microdischarges ($N_{md}$) is determined for $T_{ON}$ = 1 ms and multiplied by the duty cycle to always consider the same residence. Whatever the plasma power, $N_{md}$ is higher in pure AC mode than in burst mode. For instance, at $P_{plasma}$ = 100 W, the burst mode (for 50% duty cycle) allows the formation of 340 microdischarges per period at $P_{applied}$ = 50 W, while 480 microdischarges are obtained at $P_{applied}$ = 100 W in the pure AC mode. The burst mode presents a lower number of microdischarges, but the applied power is twice as low, which is thus more interesting in terms of energy efficiency. For the same plasma power, a lower $N_{md}$ means a higher plasma voltage ($V_{pl,eff}$), hence a higher electric field and thus higher electron temperatures. The filaments are generated in a higher electric field, leading to a higher efficiency of the $CO_2$ conversion.

Figures 10(b) and (c) give information on the microdischarge lifetime as well as the electrical charge and current versus plasma power. These electrical parameters obviously increase with plasma power and the differences between pure AC mode and burst mode are noteworthy: $Q_{pl}$ is indeed lowered in the burst mode. In terms of plasma ignition, the two modes result in a quite similar plasma charge accumulation and plasma current, when operating at the same delivered power, as observed in figure 10(c). As the plasma charge is directly related to the plasma current, the electron densities might be very similar in both modes during $T_{ON}$.

# Conclusion

We studied the $CO_2$ dissociation in a cylindrical DBD reactor operating at atmospheric pressure, focusing on the comparison between the pure AC mode and so-called burst mode, in which an AC voltage is only applied during a certain amount of time. The AC voltage signal has a period of 35.7 μs (corresponding to an applied frequency of 28.6 kHz) while the time width of the voltage burst (i.e. $T_{ON}$) is 1 ms and the $T_{OFF}$ period varies between 0 and 1.5 ms (corresponding to duty cycles ($D_{cycle}$) between 100 and 40%, respectively). Note that $D_{cycle}$ = 100% corresponds to the AC mode. Comparing this usual AC mode with the burst mode leads us to conclude that the $CO_2$ conversion is significantly higher in the burst mode, and it increases from 16.1–25.8% upon decreasing $D_{cycle}$ from 100 to 40%. Likewise, the corresponding energy efficiency of the $CO_2$ splitting rises from 14.5 to 23.1%.

To explain the impact of the plasma electrical properties on the $CO_2$ conversion, a detailed electrical characterization is carried out in the burst mode for various duty cycles. A comparison between the burst and pure AC mode at the same plasma powers reveals that the plasma currents and microdischarge mean lifetimes are quite similar, while the number of microdischarges appears lower in the burst mode.

We have also investigated the time-variation of the applied DBD voltage in the two modes: it is much more stable and constant as a function of the processing time at a low duty cycle (burst mode), whereas the voltage always drops at duty cycles higher than 70%, or in the pure AC mode. Also, we have shown that for a same plasma power, a lower $N_{md}$ means a higher plasma voltage ($V_{pl,eff}$), hence a higher electric field and thus higher electron temperatures. The filaments are generated in a higher electric field, leading to a higher efficiency of the $CO_2$ conversion.

Finally, OES measurements have shown that the burst mode leads to a lower gas temperature, indicating that a smaller fraction of the power is used for gas heating, so a larger fraction can be dedicated to the $CO_2$ dissociation. Overall we can conclude that the so-called burst mode yields a significantly higher $CO_2$ conversion for the same applied power as in a usual AC mode, which is very promising in terms of applications.

# Acknowledgments

The authors acknowledge financial support from the IAPVII/12, P7/34 (Inter-university Attraction Pole) program 'PSI-Physical Chemistry of Plasma-Surface Interactions', financially supported by the Belgian Federal Office for Science Policy (BELSPO). A. Ozkan would also like to thank financial support given by 'Fonds David et Alice Van Buuren'.